\documentclass[preprint,journal]{vgtc}            % preprint (journal style)

\onlineid{0}

\vgtccategory{Research}

\title{Read, Critique, or Sketch? Investigating Alternative Visualization Literacy Assessment Modalities}

\author{%
\authororcid{Zach Cutler}{0000-0002-2656-3413},
\authororcid{Lily W. Ge}{0000-0003-2350-8686},
\authororcid{Matthew Kay}{0000-0001-9446-0419},
\authororcid{Lane Harrison}{0000-0003-3029-2799},
\authororcid{Andrew McNutt}{0000-0001-8255-4258}, and
\authororcid{Alexander Lex}{0000-0001-6930-5468}
}

\authorfooter{
\item Zach Cutler and Andrew McNutt are with University of Utah. Email: \{u1025642, andrew.mcnutt\}@utah.edu
\item Lily Ge and Matthew Kay are with Northwestern University. Email: \{wanqian.ge, mjskay\}@northwestern.edu
\item Lane Harrison is with WPI. Email: ltharrison@wpi.edu
\item A. Lex is with Graz University of Technology and the University of Utah. E-Mail: alex@visdesignlab.net
}

\abstract{%
Visualization literacy is a multifaceted construct encompassing skills and competencies, such as decoding data, constructing charts, and identifying design flaws. Yet, assessments of these competencies has been primarily constrained to multiple choice assessments that target lower-order skills, such as chart comprehension. As a result, they often exhibit ceiling effects (i.e., even modestly skilled individuals commonly score near the top of the scale), and do not provide enough information about an individual's higher-order skills (e.g., applying external knowledge, formulating critiques, and designing visualizations). 
To close these gaps, we develop and investigate two web-based qualitative assessments for testing the critique and design aspects of visualization literacy through online think-aloud critique and sketching of visualization designs based on data and a prompt. 
We compare performance on our assessments to two established visualization literacy assessments, CALVI and Mini-VLAT, by administering them to three groups that represent three experience levels: crowdworkers, students who have taken a relevant course, and researchers. 
We find that our critique and sketching assessments capture skills distinct from existing measures and that they differentiate between experienced individuals better than multiple choice-based alternatives. Although administering and grading qualitative assessments can be challenging, our findings suggest qualitative, multimodal assessments are a promising complement to existing visualization literacy assessments, in particular when high visualization skills need to be distinguished. 
}

\keywords{Visualization Literacy, Critique, Sketch, Think Aloud}

\usepackage{xcolor}
\usepackage{soul}
\newcommand{\etal}{et al.}
\newcommand{\etals}{et al.'s}

\newcommand{\eg}{{e.g.}}

\newcommand\am[1]{{\color{red}[AM: #1]}}

\newcommand\revised[1]{{#1}}

\usepackage{xspace}
\usepackage[pagebackref,bookmarks]{hyperref}

\definecolor{steelblue}{RGB}{70, 130, 180}

\hypersetup{
    colorlinks=true, 
    linkcolor=steelblue, 
    urlcolor=steelblue,
    citecolor=steelblue
}

\let\oldhref\href

\newcommand{\ourhref}[2]{\oldhref{#1}{\textcolor{steelblue}{$\nearrow$ #2}}}

\newcommand{\secref}[1]{\hyperref[#1]{Sec.~\ref*{#1}}}
\newcommand{\appendixref}[1]{\hyperref[#1]{Appendix~\ref*{#1}}}
\newcommand{\figref}[1]{\hyperref[#1]{Fig.~\ref*{#1}}}
\newcommand{\eqnref}[1]{\hyperref[#1]{Eqn.~\ref*{#1}}}
\newcommand{\tabref}[1]{\hyperref[#1]{Table ~\ref*{#1}}}

\definecolor{quoteColor}{HTML}{ff5733}
\newcommand\qt[1]{\emph{``#1''}}
\newcommand{\pxx}[1]
{\textbf{P}$_{\textrm{#1}}$}

\newcommand{\citeauthor}[1]{\am{TODO: #1, \cite{#1}}}
\newcommand{\citet}[1]{\citeauthor{#1}}

\newcommand{\paraheadd}[1]
{%
    \vspace{0.07in}%
    \noindent%
    \textbf{\textit{#1}}%
}
\newcommand{\parahead}[1]{\paraheadd{#1.}}
\def\subsubsec#1
{\subsubsection{#1}}

\setuldepth{Berlin}
\definecolor{linkColor}{HTML}{257E98}

\definecolor{synthesisColor}{HTML}{f0f0f0}

\definecolor{takeawayColor}{HTML}{F9EFCE}
\definecolor{takeawayAccent}{HTML}{DFB11B}
\usepackage{multirow}

\usepackage[most]{tcolorbox}
\tcbuselibrary{skins,breakable}
\newtcolorbox{mybox}[2][]{breakable,sharp corners, skin=enhancedmiddle jigsaw,parbox=false,
boxrule=0mm,leftrule=2mm,boxsep=0mm,arc=0mm,outer arc=0mm,attach title to upper,
after title={.\ }, coltitle=black,colback=takeawayColor,colframe=takeawayAccent,
fonttitle=\bfseries,#1}

\teaser{
  \centering
    \includegraphics[width=\linewidth]{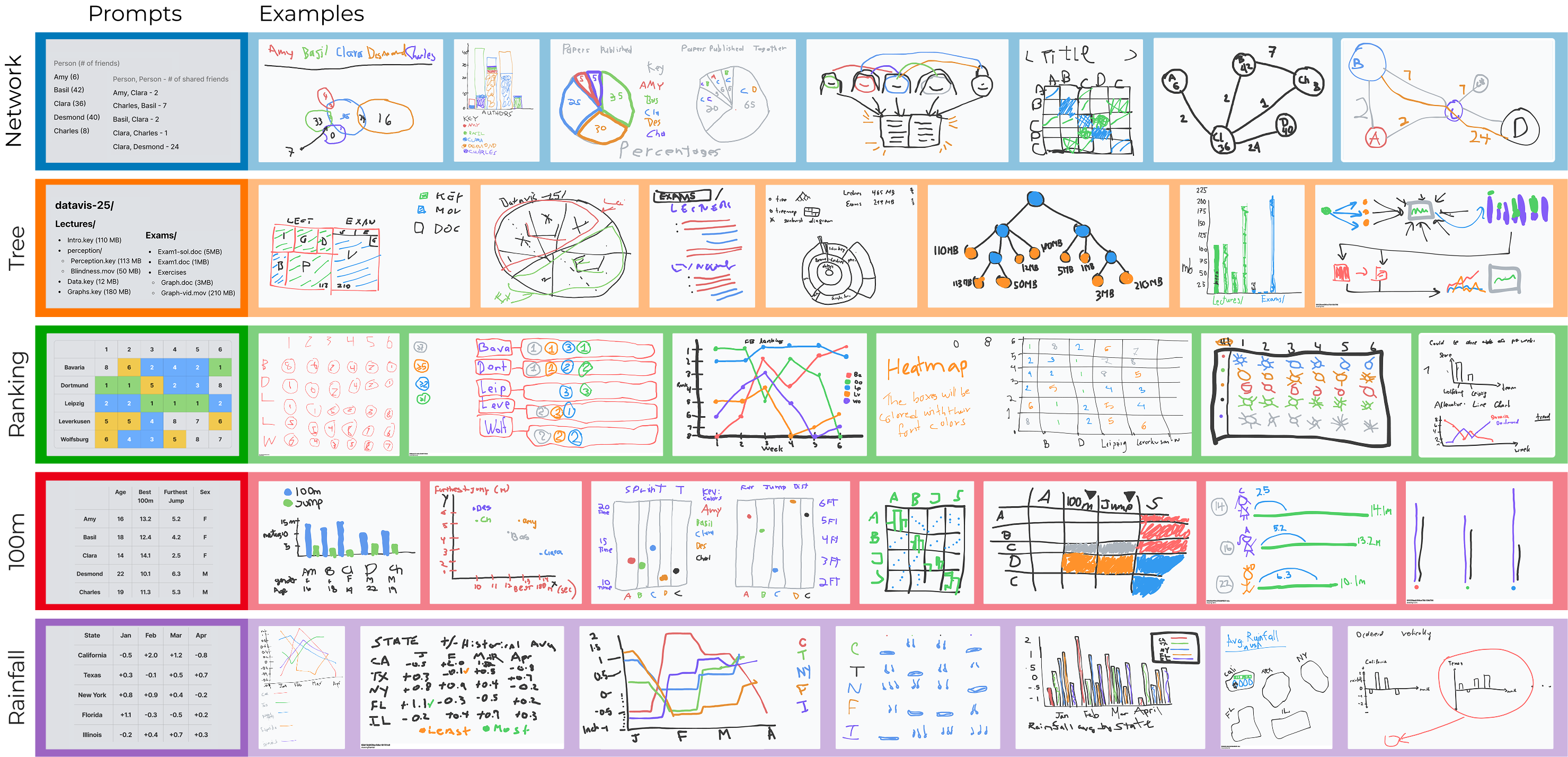}
    \caption{Examples of the drawings that participants generated in different sketching tasks, based on the data snippets shown on the left. In each task participants were asked to produce the best visualization for the given data. Participants produced a wide range of visualization types and encodings for each question.}
    \label{fig:drawing-comics}
}

\graphicspath{{figs/}{figures/}{pictures/}{images/}{./}} % where to search for the images

\usepackage{tabu}                      % only used for the table example
\usepackage{booktabs}                  % only used for the table example
\usepackage{lipsum}                    % used to generate placeholder text
\usepackage{mwe}                       % used to generate placeholder figures
\usepackage{ccicons}                   % package to be able to use icons from creative commons
\usepackage[table]{xcolor}
\usepackage{colortbl}

\usepackage{mathptmx}                  % use matching math font

\begin{document}

%%%%%%%%%%%%%%%%%%%%%%%%%%%%%%%%%%%%%%%%%%%%%%%%%%%%%%%%%%%%%%%%
%%%%%%%%%%%%%%%%%%%%%% START OF THE PAPER %%%%%%%%%%%%%%%%%%%%%%
%%%%%%%%%%%%%%%%%%%%%%%%%%%%%%%%%%%%%%%%%%%%%%%%%%%%%%%%%%%%%%%%

%% The ``\maketitle'' command must be the first command after the
%% ``\begin{document}'' command. It prepares and prints the title block.
%% the only exception to this rule is the \firstsection command

\maketitle

\firstsection{Introduction}
\section{Introduction}

To design visualizations for a target audience, designers need to know what kinds of visualizations their audience understands. Similarly, to reliably interpret the results of a study about a visualization technique, researchers need to accurately judge participants' abilities. The visualization community has recognized these needs and has turned to measuring visualization literacy to answer these questions.  
However, visualization literacy is complex, involving numerous skills and competencies~\cite{leon_multiliteracy_2026, varona_state_2025, hedayati_pixels_2024} for both reading visualizations and constructing them. 
%  (or graphicacy as it was initially known)
Early visualization literacy assessments focused on reading via both high-level skills, such as visualization comprehension~\cite{varona_state_2025}, and lower-level ones, such as the ability to extract values from charts or identify trends. 
Tests such as the Visualization Literacy Assessment Test (VLAT) \cite{lee_vlat_2017} and the short-form Mini-VLAT\cite{pandey_Mini-VLAT_2023} ask participants to answer factual questions about data depicted in a visualization. 
More recently, additional aspects of visualization literacy have been identified, and some have received their own assessments, such as critical thinking
(Critical Thinking Assessment for Literacy in Visualizations, CALVI~\cite{ge_calvi_2023}), 
and visualization construction (Assessment of Visual Encoding Ability in Visualization Construction, AVEC~\cite{ge_avec_2025}).

However, most existing visualization literacy assessments are multiple-choice and focus on basic visualization reading skills. While quantitative assessments allow for rapid, unambiguous grading of right and wrong answers, they lack the nuance of qualitative assessments. 
Consider an analogy:
% Despite the drawbacks, 
qualitative, unconstrained, creation-focused assessments for textual literacy are commonplace, and are included in institutionalized evaluations \revised{such as the SAT and GRE, which serve as the primary standardized tests for entrance into American undergraduate and graduate programs, respectively.
% ---both of which feature qualitatively graded writing components. 
In a report to the College Board, which develops and administers the SAT,} Breland \etal{}~\cite{breland_writing_1999} argue that, despite their drawbacks, written essays are an indispensable part of written skills assessments, observing \qt{Multiple-choice tests tend to assess the ability to choose the revision or revision strategy that could improve the coherence or correctness of text. Essay tests, on the other hand, tend to assess the ability of examinees to reflect on a topic and then engage in the process of conceiving, synthesizing, and articulating their own thoughts about the topic.} 
Both Breland~\cite{breland_writing_1999} and others~\cite{cooper_assessment_1984} argue a combination of multiple choice and qualitative assessments more fully evaluates the range of skills involved in text literacy. 

The visualization community, however, has not seriously explored unconstrained and creation-focused qualitative assessments of visualization literacy~\cite{ge_autoethnography_2026}. Current assessments are unable to evaluate higher level visualization comprehension skills (\eg{} design empathy~\cite{hedayati_what_2025} or application of external knowledge) or visualization construction skills (\eg{} applying data transformations or use of visual hierarchy). \revised{In a recent auto-ethnography of visualization literacy researchers, Ge \etal{}~\cite{ge_autoethnography_2026} recognize the constrained set of skills that current assessment modalities measure, and call for utilizing qualitative methods that may be more suited to measuring higher-order skills than existing assessments.}
% Further, existing assessments are strongly subject to a ceiling effect, in which \am{participants} of differing skill levels can not be differentiated because they have identically perfect scores. 
To address this gap, we contribute \textbf{two qualitative visualization literacy assessments}, investigating \textbf{visualization construction through \textbf{Sketching}} via freeform sketching, and \textbf{visualization comprehension through \textbf{Critique}} via think-aloud.
We focus on these evaluations because they are common-place parts of academic visualization curricula~\cite{hedayati_what_2025}, and afford substantial elaborative nuance. 
We explore how these assessments compare with prior tests through an (n=80) online user study consisting of both novel assessments, as well as CALVI and Mini-VLAT.
% ---in order to compare the relative  strength of each of the assessments.
We recruit three groups with ostensibly distinct skill levels,  crowd workers, students who have taken a visualization course, and visualization researchers. 

We find that \textbf{both sketching and critique succeed at differentiating between groups of varying skill level and exhibit improved sensitivity} over both Mini-VLAT and CALVI. 
Correlation analysis shows that \textbf{our qualitative assessments capture distinct skills} from existing assessments, suggesting that sketching and critique measures aspects of visualization literacy which CALVI and Mini-VLAT do not. 
Our findings suggest that qualitative assessments should play an important role in future development of visualization literacy assessments, and should be considered with or in place of existing tests, particularly when evaluating individuals with visualization training.
\section{Related Work}

This work builds on prior visualization literacy research and assessments, as well as on sketching and critique generally. 

\begin{figure*}
    \centering
    \includegraphics[width=\linewidth]{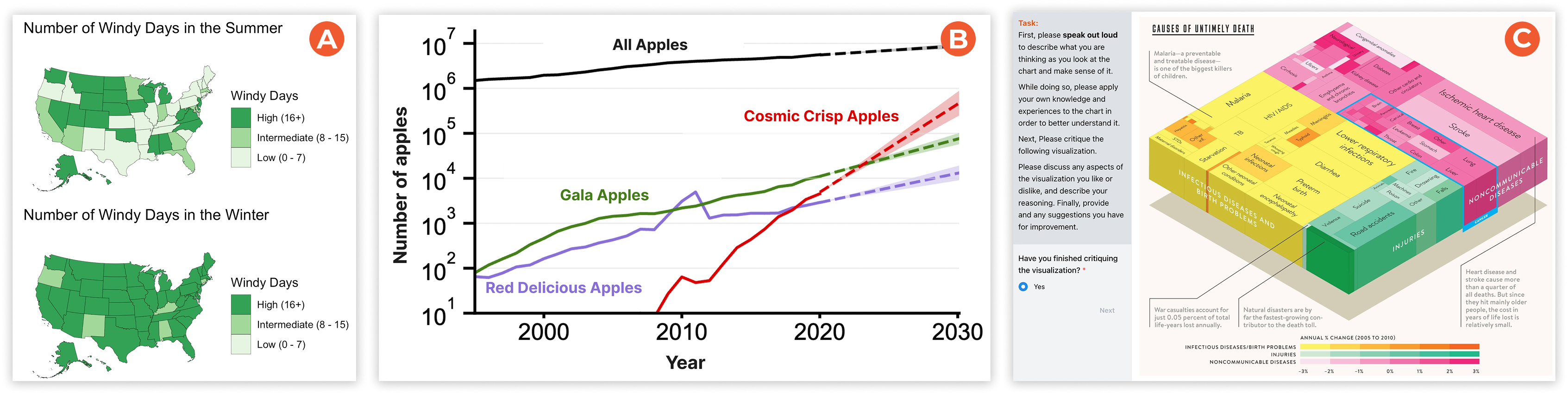}
    \caption{The visualizations used in the critique-based assessment of visualization literacy. 
    (A) A pair of maps comparing windy days in the summer to the winter, previously used in CALVI~\cite{ge_calvi_2023}. The chart uses a familiar encoding, but a badly binned color scale. 
    (B) A line chart on a logarithmic scale with seemingly made-up projections. The original chart shows the rise of paper mill publications~\cite{richardson_entities_2025}, which we modified to apple varieties.
    (C) A 3D treemap of ``Causes of Untimely Death'' originally published in Wired and made popular by Bill Gates. The chart type is complex, as is the dataset. 
    % am: reordered to match the complexity order from the text
    }
    \label{fig:crit-images}
        \vspace{-2em}
\end{figure*}

\revised{\subsection{Visualization Literacy}}

While visualization literacy has generally been thought of as the ability to read and create visualizations, it has been defined in a variety of ways~\cite{lee_vlat_2017, ge_calvi_2023}. The continuous efforts demonstrate the difficulties in precisely defining visualization literacy, which has led to challenges operationalizing it in research studies. Ge \etal{}~\cite{ge_autoethnography_2026} unpack and discuss the general difficulty of defining visualization literacy, stressing that it is domain-specific, culturally dependent, and evolves over time.

Despite its definitional difficulties, many attempts have been made to categorize the skills associated with visualization literacy. Varona \etal{}~\cite{varona_state_2025} classify visualization literacy tests into four contexts, \textit{consumption}, \textit{construction}, \textit{critique}, and \textit{connection}. Hedayati \etal{}~\cite{hedayati_pixels_2024} describe visualization literacy as having three main components, \textit{visualization competencies}, \textit{visualization comprehension processes}, and \textit{visualization practices}. 
Leon \etal{}~\cite{leon_multiliteracy_2026} describe interactive visualization literacy and define three levels of abstraction that each contain separate sub-literacies, such as \textit{operational} (\eg{} tool literacy), \textit{tactical} (\eg{} task literacy, interaction literacy), and \textit{strategic} (\eg{} workflow literacy, insight literacy). While we do not subscribe to any of these models specifically, we borrow from each in developing our assessment rubrics and for identifying the precise skills that we set out to measure. 

Cope and Kalatzis~\cite {cope_things} describe the theory of \textit{multiliteracies}. Multiliteracies explains that literacy can be multimodal, expanding beyond classical definitions of literacy focused on text. 
This has led to acknowledgment of the relationship between (and plurality of) literacies~\cite{leon_multiliteracy_2026} in visualization. 
% Visualization literacy is a prime example of this entanglement, as 
The skills required for high visualization literacy are directly related to other competencies, including data, media, and statistics. 
Both of our assessments (especially critique) directly require data literacy skills, which are included as a part of our rubric.

\revised{\subsection{Visualization Literacy Measures}} 
A range of works have focused on developing measures for visualization literacy. Boy \etal{}~\cite{boy_principled_2014} proposed a way of measuring a person's visualization literacy utilizing Item Response Theory~\cite{demars_item_2010} to evaluate assessments, a practice that is now standard and we also utilize for our own assessments. Lee \etal{}~\cite{lee_vlat_2017}  later introduced the widely adopted Visualization Literacy Assessment Test (VLAT), which has been followed by variations such as the Mini-VLAT by Pandey \etal{}~\cite{pandey_Mini-VLAT_2023}, and the adaptive VLAT by Cui \etal{}~\cite{cui_adaptive_2024}.
VLAT and its derivatives measure a person's visualization comprehension, or how well they can extract information from a chart. 
% \am{feel like SGL~\cite{okan_using_2019} should be described up here}

Other literacy assessments target different visualization literacy skills. 
CALVI~\cite{ge_calvi_2023} and A-CALVI~\cite{cui_adaptive_2024} measure critical thinking skills by asking participants to identify design errors commonly used to misinform, such as inverted axes and poorly selected color scales. 
AVEC~\cite{ge_avec_2025} and Iguanadon~\cite{adelberger_iguanodon_2025} measure visualization construction literacy. 
Unique among assessments, AVEC tasks participants to create appropriate visualizations for a task by mapping data to appropriate visual channels, such as position, color, and size, measuring skills that may be useful when using common visualization tools such as Tableau or Excel.
However, AVEC does not measure freeform construction and only allows a limited expressive range. 
Moreover, none of these assessments use unconstrained qualitative methods, limiting the skill types they assess. 

Most closely related to our work is Multidimensional Assessment Method for  Visualization Understanding (MdamV)~\cite{saske_multidimensional_2026} which targets skills beyond comprehension, including \textit{critique} and \textit{contextualization} (an item in our critique rubric)---overlapping our assessments. 
However, like other assessments, MdamV primarily relies on multiple choice questions and self-assessed scales, with the exception of optional text boxes for critique that less than half of participants utilized.  

Chang \etal~\cite{chang_tell_2026} investigate the correlation between three different literacy assessments, Mini-VLAT, CALVI, and a self-assessment, SGL~\cite{okan_using_2019}. 
They find that, while somewhat correlated, the three assessments measure distinct skills. We conduct a similar analysis on our own assessments as well as Mini-VLAT and CALVI.

\vspace{1em}
\revised{\subsection{Sketching \& Critique}}
% \zc{stopped here}
% * Intro sentence
 Many skills involved in reading and creating visualizations cannot be fully captured by selected-response formats. Assessment format heavily influences the skills an assessment can capture~\cite{ge_autoethnography_2026}, leading us to opt for formats better suited to open-ended tasks, such as freeform sketching and open-ended critique.

% * Sketching intro. 
Sketching has been used as a proxy for visualization creation in prior studies in the visualization community. Walny \etal{}~\cite{walny_exploratory_2015} conduct a qualitative user study asking participants to sketch data visualizations by hand, not unlike our own assessment. Their findings influenced our own rubric, such as their observation of the importance of external knowledge to sketching. As the focus of their sketching study was on qualitative learnings, they did not create rubrics or grade any of the sketches. Roberts \etal{}~\cite{roberts_sketching_2016} discuss the use of sketching as a way to plan visualization design, and describe how sketching encourages divergent solutions. Other fields have investigated sketching as a means of assessment. Merzdorf \etal{}~\cite{merzdorf_sketching_2024}, conduct a literature review of sketching assessments, and find a number of assessments developed on the fields of Engineering, Design, and Art. Sketching is also a common means of assessment in visualization courses\cite{hedayati_what_2025}. However, visualization literacy assessments have not yet utilized sketching as an assessment method, and we are not aware of any rubrics for evaluating visualization design sketches.

Critique of data visualizations is also a skill commonly taught in visualization courses~\cite{hedayati_what_2025}, though more formally trained and evaluated as a skill in other fields, such as Art, which have developed rubrics for assessing formal critiques~\cite{blommel_rubric_2007, tam_evaluating_2018} that we borrow heavily from. Rubrics for critique frequently emphasize similar skills as our own, such as contextualizing, reasoning, and interpretation. None of these rubrics are for assessing open-ended critiques of visualizations, which involves its own unique set of skills to be considered in a rubric, such as design empathy~\cite{hedayati_what_2025}. Visualization researchers have also considered critique as a form of evaluation~\cite{kosara_visualization_2007, brath_evaluation_2016}. Roberts \etal{}~\cite{roberts_critical_2026} develop a heuristic-based approach to critiquing visualizations, based on a series of low-level heuristics (\eg{}, "Would be useful", and "Aesthetic and visually attractive") graded 1-5. However, critique as a form of investigating visualization literacy has not been investigated.
\vspace{1em}
\section{Measuring Visualization Construction and Critique}
\label{sec:measuring}

To investigate the feasibility of qualitative assessments targeting higher-order skills we develop two new assessments, \textbf{Sketching} and \textbf{Critique}. We were influenced by Varona \etal{}~\cite{varona_state_2025}, who describe four competency themes for visualization literacy, including skills beyond classical visualization comprehension, such as \textit{construction} and \textit{critique}. Although assessments exist that target both skills, such as CALVI for critique and AVEC for construction, their modalities limit the skills that they can assess. We develop assessments that build on these past tests by introducing qualitative modalities that are suitable for targeting different, higher-order skills.

% \subsection{Assessments}

% \am{intro}

\parahead{Sketching}
In the sketching assessment, participants were given five datasets and prompted to visualize them. The datasets were created by a co-author for the purpose of in-class exercises, where students were asked to explore different encodings for datasets before visualizations were formally introduced. They emulate real-world datasets, but are designed such that visualizations can be easily sketched even when all data elements are included. The datasets partially cover the dataset types and attribute types as described by Munzner~\cite{munzner_visualization_2014}. The datasets, shown in \figref{fig:drawing-comics}, are a social network, a file hierarchy (tree), a table with rankings, a heterogeneous table with athletic competition results and athlete characteristics, and a homogeneous table with rainfall data. Attributes types include categorical data, ordered data, quantitative data, sequential data, and diverging data. None of the datasets are geometric, due to the expected difficulty of converting such data to sketch by hand.

% \al{@Zach Cutler you're treating rainfall data as fields as by Munzner, implicitly; I don't understand why.
% I mean, in the furthest sense, sure, but at least usually when we talk about visualizing fields you don't just pick 5 things and show them on a table, but you visualize it somewhat continuously, so I'd argue we cover only tables and networks, and for that we mgiht even drop the munzner reference}

\parahead{Critique}
In the critique-based assessment, participants are shown a series of visualizations from \figref{fig:crit-images} in a random order. We selected three visualizations of different difficulties based on the complexity of the data and popular familiarity of the chart types, with the Map visualization being the easiest (common representation, simple dataset), the treemap being the most difficult (uncommon representation, complex dataset), and the linechart in the middle (challenging data properties). 

We chose visualizations that are not ``perfect'' so they would be amenable to critique and suggestions for improvement. They all contain a range of problems which require various types of knowledge and skills related to visualization literacy to properly critique. 

For example, the \textbf{treemap} visualization has a sequential color scale which ranges from -3 to 3, with a zero in the middle, where a diverging color scale would be more appropriate. It also uses an unfamiliar, space-filling encoding and a 3D effect. The chart was popularized by the Washington Post~\cite{wonkborg_bill_2013}, citing it as Bill Gates' favorite chart of 2013. The visualization was widely~\cite{few_one_2014, gelman_bill_2013} critiqued (and redesigned) and has been a staple of various data visualization courses.

The \textbf{line chart} has the problem of making linear projections seemingly based only on the slope of the known data. This problem is amplified given that the chart has a log scale, resulting in an exponential projection. The original chart, from Richardson \etal{}~\cite{richardson_entities_2025}, displayed the rise of scientific publication mills. The visualization was shared on LinkedIn without context explaining the dubious projection, leading to widespread critique of the chart.
We modified the chart's domain because we considered scientific publications to not be an understandable topic for a general audience, and replaced the line labels with apple varieties, highlighting the rise of ``Cosmic Crisp Apples''. Recognizing the issues with the chart requires applying a mix of data literacy, critical thinking, and visualization design skills to identify.

The \textbf{map} is an intentionally misleading visualization originally created for CALVI~\cite{ge_calvi_2023}. It features a poor selection of bins for the color scale, resulting in the misleading appearance that much of the country in the winter has the same number of windy days. Actual values are not retrievable from the chart. Again, data literacy, critical thinking, and visualization design knowledge play a role in identifying this problem.

While these are the most prominent issues in each chart, each chart has many aspects that resulted in broad discussion, praise, and critique. 

\subsection{Rubric Development}
\label{sec:rubrics}

A key part of our contribution is the grading scheme for the multimodal data generated in the study, which consisted of (for sketching) digital sketches accompanied by spoken explanations and (for critique) verbal responses.

Details on the procedure are described in \autoref{sec:methods}. 
For grading such open-ended responses, there are two types of approaches. \textit{Analytic scoring} specifies features of a response and assigns each feature a point value. In contrast, \textit{holistic scoring} involves graders making judgments about the quality of a response based on a variety of influences~\cite{livingston_constructed-response_2009}. In early pilots, we took an analytic approach to grading. For critique, we listed all important aspects of each visualization we expected to hear from participants (\eg{} poor continuous color encoding and the 3D aspect of the treemap), quantified how good or bad each was, and how important each aspect was to understanding the visualization. We scored responses based on how much of the rubric participants covered. For sketching, we listed the aspects of the data we believed important to encode (\eg{} hierarchy and size in the tree dataset) of each dataset, and assigned them point values based on how suitable their chosen encoding was based on visualization guidelines.

To test our grading approach, we piloted with five crowd workers from Prolific. The first two authors graded the responses using this rubric and then met with the broader team to iterate. However, we immediately felt that this approach was not adequate for capturing the diverse ways that participants could demonstrate literacy, and also did not allow for the subjective nature of visualization design. For example, the 3D aspect of the treemap is critiqued by many participants, frequently citing common visualization practice, readability, and perceptual concerns. Other participants noted that the 3D effect allows for creative labeling of the treemaps categories, minimal perceptual concerns due to the isometric design, and a more aesthetic visualization which is ostensibly designed for public consumption. In such a judgment case, which side of the 3D argument a participant lands on says little about their visualization literacy. Instead, the rationale and reasoning expressed to justify their arguments are what demonstrate expertise.
To address these shortcomings of the rubric, we looked for more holistic ways to evaluate visualization designs and critique. 

\parahead{Sketching Rubric} While rubrics have been developed for grading visualization projects created in courses~\cite{beasley_leveraging_2020, friedman_leveraging_2021}, many are aimed at completed, fully engineered projects, not initial design sketches. 
\revised{Instead, we leveraged co-authors' experience grading design sketches for course examinations to develop the rubric that measures four distinct skills involved in the visualization creation process. Hence, the sketching rubric has four categories: \textit{Visualization Design / Visual Hierarchy}, \textit{Base Encoding Choice}, \textit{Semantic Validity}, and \textit{Guides/Text/Annotations}.} Each of these categories is graded on a five-point scale, ranging from \textit{None} to \textit{Excellent}. More details on the rubric can be found in \autoref{tab:compressed-rubric}.

Initially, we intended to grade sketching by listening to and observing the entire process that went into creating the final visualization. However, we observed far fewer process-related think-alouds than we expected, possibly due to the already high cognitive load of designing visualizations. While the replays contained rich data about the process a participant went through to create a final sketch, it was not clear how to incorporate this information into a final grade. 
As such, we graded sketches based only on final results and the transcribed audio data. 

Particularly, for ambiguous or less detailed sketches, we used transcripts to provide context.
For example, many sketches do not have clear axis labels or legends, but the corresponding audio indicates that such details are implied. In this situation, we treated the sketch as if it included any features described by the participant. We do not judge participants' artistic ability or the aesthetics of final sketches, and instead focus on the design choices that went into the sketches. In rare cases when transcribed audio data and sketches were not clear, we examined the full replay. Although the replay process was not considered in scoring, we believe it worthy of further examination (\secref{limitations-future-work}).

\parahead{Critique Rubric} To generate our rubric for the critique-based assessment, we borrow heavily from arts education, which has placed considerable focus on developing means to teach and evaluate critique as a skill. We specifically were influenced by the work of Tam~\cite{tam_evaluating_2018}, and all of our rubric categories resemble one or multiple categories in their work. Our rubric consists of three categories, \textit{Reading}, \textit{Contextualization}, and \textit{Judgment} (see \autoref{tab:compressed-rubric}). Again, the categories are graded on a five-point scale, ranging from None to Excellent. These rubric categories reflect the broad set of skills that go into critiquing a chart, and can loosely be mapped to Varona \etals{}~\cite{varona_state_2025} categories of consumption (Reading), connection (Contextualization) and critique (Judgment). 
We initially explored developing bespoke rubrics for each chart, however we converged on a single rubric as it would better support reuse. Practical application of the rubric, however, does change from question to question, depending on important aspects of the chart. For example, application of \textit{contextualization}  shifts depending on what skills are necessary to critique a chart. For our \textbf{line chart} question, data literacy and critical thinking are primary drivers of contextualization. For the \textbf{treemap}, design empathy is critical to contextualization.

\begin{table}[tb]
\centering
\footnotesize
    \begin{tabular}{@{}c >{\raggedright\arraybackslash}p{0.9\linewidth}@{}}
        \toprule
        \multirow[b]{2.5}{*}{\rotatebox[origin=c]{90}{\textbf{Sketching}}}
        & \textbf{Visualization Design / Visual Hierarchy} \newline
        Visual organization and layout creates structure and intentionally guides attention \\
        \addlinespace
        & \textbf{Semantic Validity} \newline
        Visual encodings allow for underlying trends and relationships in the data to be found \\
        \addlinespace
        & \textbf{Guides, Annotations, Text} \newline
        Labeling, such as axes, legends, and annotations, are well chosen and designed \\
        \addlinespace
        & \textbf{Base Encoding} \newline
        The primary visual encodings are well chosen for the data \& primary encoding is will suited to the data \\
        \midrule
        \multirow[b]{2.3}{*}{\rotatebox[origin=c]{90}{\textbf{Critique}}}
        
        & \textbf{Reading} \newline
        Understanding the content and encodings of a visualization. \\
        \addlinespace
        & \textbf{Contextualization} \newline
        Applying other knowledge (e.g., data literacy or external knowledge) to understand and critique the chart \\
        \addlinespace
        & \textbf{Judgment} \newline
        Identifying and reasoning about in/effective visualization components \\
        \bottomrule
    \end{tabular}
    \caption{We developed rubrics for both sketching and critiquing activities, here shown by the high-level capability that each rubric item tested. Each capability was scored on as Excellent, Good, Satisfactory, Poor, or None. Full rubrics are available in supplemental material} 
    \label{tab:compressed-rubric}
    \vspace{-2em}
\end{table}

\begin{figure}[t]
    \centering
    \includegraphics[width=\linewidth]{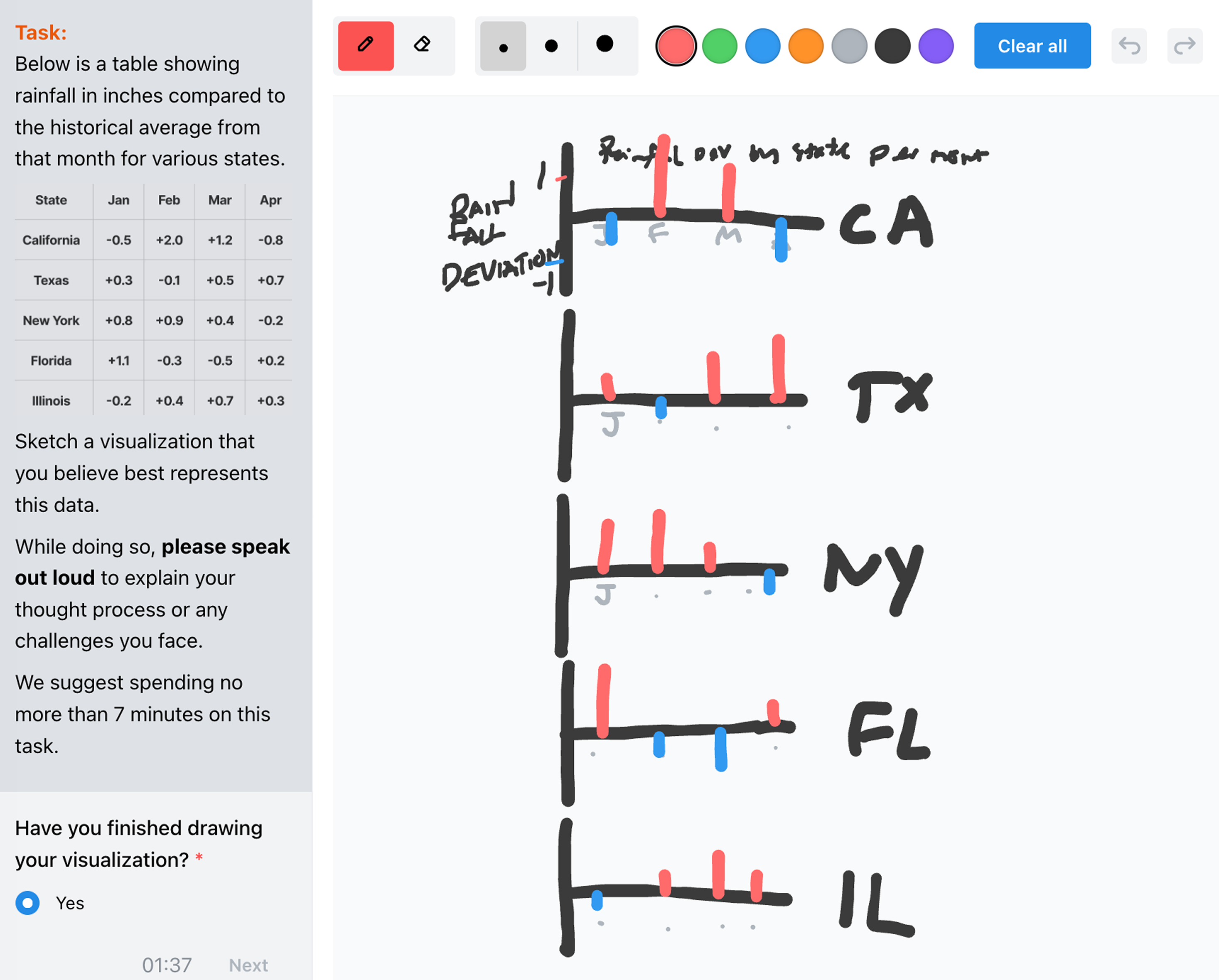}
    \caption{Prompt and canvas participants saw during our sketching tasks, showing the rainfall dataset and a participants solution. In addition to standard drawing functions like pen size or erase, participants were able to pan and zoom---facilitating larger or more detailed sketches.}
    \label{fig:drawingInterface}
    \vspace{-2em}
\end{figure}

\section{Study Methods}
\label{sec:methods}

To evaluate our assessments, we conduct a user study that we design and analyze based on three primary research questions.

\subsection{Research Questions}
\newcommand{\rqq}[1]{\paraheadd{#1}}

\rqq{Will Sketching and Critique assessments be able to differentiate between groups with varying visualization expertise?}\\
The primary motivation of this work is to develop assessments targeting higher-order skills that can only be revealed through qualitative assessments. Assessing higher-order skills should differentiate between even experienced individuals, avoiding ceiling effects that have been observed in VLAT. To test if Sketching and Critique achieve this goal, we conduct a user study on participants with varying levels of expertise. 

\rqq{Will Sketching and Critique assessments be able to discriminate participants with different skills?}\\
Several previous visualization literacy assessments~\cite{ge_calvi_2023, boy_principled_2014, ge_avec_2025, pandey_Mini-VLAT_2023} use Bayesian Item Response Theory (IRT) to evaluate their tests, including frequently reporting on \textit{item discrimination}. Item discrimination describes a question's ability to differentiate participants with different abilities. Acceptable item discrimination would indicate that our assessments do not only differentiate between our three participant groups, but also can discriminate between participants within each group. With this goal in mind, we use Bayesian IRT to evaluate our assessments.

\rqq{How correlated are the Sketching and Critique assessments with each other and with CALVI/Mini-VLAT?} \\
Lastly, we seek to understand the relationship between our assessments and prior methods of assessing visualization literacy. 
% To compare our assessments to established measures of visualization literacy, 
Our within-subjects user study includes CALVI and Mini-VLAT, allowing us to compare performance on all four assessments. 
Understanding the correlation between tests is important for choosing a test. For example, if understanding visualization literacy with the Mini-VLAT perfectly correlates with our critique scale, it would be more efficient to use the comparatively simpler multiple-choice tests to measure reading and critique performance. 
We utilize our IRT models to show population-level correlation estimates, as well as investigate individual variation in participants' performance.

\subsection{Study Procedure}

To explore these research questions, we conducted a within-subjects/mixed subjects study design consisting of our two assessments, Sketching and Critique, as well as a combination of CALVI and Mini-VLAT questions. 
The study was implemented with reVISit~\cite{cutler_revisit_2026} and is available at  \ourhref{https://vdl.sci.utah.edu/visLiteracyStudy/literacy-prolific}{vdl.sci.utah.edu/visLiteracyStudy/literacy-prolific}. The source code of the study is available on \ourhref{https://github.com/visdesignlab/visLiteracyStudy}{github}.

In each task in the \textbf{Sketching} assessment, participants were shown a short description of a dataset, after which they were asked to ``Sketch a visualization that you believe best represents this data'' (see \figref{fig:drawingInterface}). 
Participants were asked to think aloud while completing the task, and encouraged not to spend more than seven minutes on each individual task, although there was no time limit. 
There was no minimum time or interaction required. Sketching is implemented as a TLDraw~\cite{tldraw_tldraw_2026} canvas with provenance tracking using Trrack~\cite{cutler_trrack_2020}. 

In each task in the \textbf{Critique} assessment, participants were shown an image of a visualization and asked to make sense of it and critique it out loud (see \figref{fig:crit-images}C). There was no time minimum or maximum, participants could continue at any time. 

Lastly, the \textbf{CALVI + Mini-VLAT} section combined CALVI and Mini-VLAT to reduce the time required to complete both assessments, as first used by Chang \etal{}~\cite{chang_tell_2026}. The 12 questions in Mini-VLAT~\cite{pandey_Mini-VLAT_2023} and the 15 ``trick'' questions in CALVI~\cite{ge_calvi_2023} are combined by replacing the ``normal'' questions from CALVI (which serve as distractors in the CALVI test) with Mini-VLAT questions. 
The order of the resulting 27 questions is random. 
We chose CALVI and Mini-VLAT because they are two of the most commonly used visualization literacy assessments. We also considered other assessments, such as AVEC and the adaptive version of CALVI and VLAT~\cite{cui_adaptive_2024}, but were concerned about adding time-consuming assessments to our own lengthy assessments.

\begin{figure}[t]
    \centering
    \includegraphics[width=\linewidth]{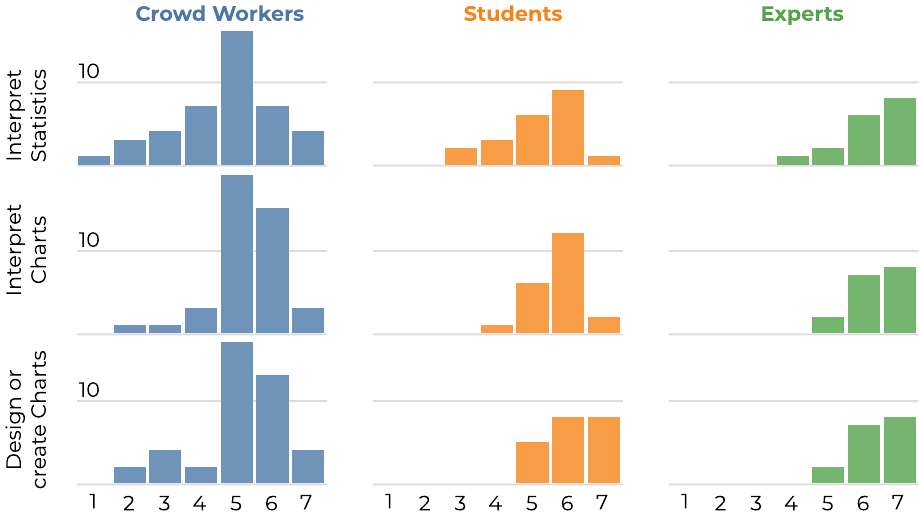}
    \caption{\textbf{Distribution of self-assessed skills for interpreting statistics and charts, as well as designing charts, increases with expertise level of participant groups.} Responses to self-assessed visualization and statistical expertise by participant group. }
    \label{fig:res_self-assesed-vis-skills}
    \vspace{-2em}
\end{figure}

To avoid biasing participants toward visualization types shown in other sections, the sketching section is always shown first, then critique, and finally CALVI/Mini-VLAT. 
The order of questions within each section is randomized. After completion of the main tasks, we asked a series of demographic and experiential questions regarding the study. 
\figref{fig:res_self-assesed-vis-skills} shows the self-assessed visualization skills in interpreting graphs, designing graphs, and interpreting numbers between groups. 

 This study was marked exempt by the University of Utah IRB (IRB\#00193994).
 % from full review.
We conducted a number of pilots, primarily to evaluate and iterate on rubrics, as described in \secref{sec:rubrics}. Initially, we intended to run our study on tablets to improve the sketching experience.
% that touchscreens offer over a mouse. 
However, early pilots revealed that it is prohibitively difficult to recruit participants with tablets.
Later pilots (and the final study) allowed for either tablet or desktop.

\parahead{Participants}
As a goal of our work is to differentiate between expertise levels, we recruited participants with a (presumed) range of experience levels by drawing from three distinct groups. 
First, we recruited \textbf{crowd workers} from Prolific, who are likely to include many individuals without formal visualization training. We filtered the participant pool to include only participants from the US/UK/Canada with a $\geq$95\% approval rating to avoid transcription difficulties with non-native English speakers, following other crowd work think-aloud studies~\cite{cutler_crowdsourced_2025}. 

We also recruited \textbf{students} who had taken a university (undergraduate or graduate) course in visualization, via departmental Slack channels as well as by advertisements distributed by visualization course instructors to their former students. These participants received formal visualization training, but are not visualization professionals.

Finally, we recruited visualization \textbf{researchers}, defined as anyone with or working towards a PhD in visualization, or who could reasonably review a visualization research paper.
We recruited researchers via personal contacts and advertising on shared research Slack channels. 

We recruited 17 experts (11 identified as men, 6 as women, aged 18-65+), 21 students (15 identified as men, 5 as women, 1 as genderfluid / gender non-conforming, aged 21-34), and 50 crowdworkers (22 identified as men, 19 as women, 1 as questioning, aged 18-54 in our final sample) as participants. Two crowdworkers were removed from analysis for leaving a majority of questions blank. Two were removed due to suspicion of AI use, due to non-aligned audio and sketches (\eg{} said they were drawing a treemap while drawing a bar chart) and sparse audio only at the beginning or end of the task. Due to a technical issue, Calvi/Mini-VLAT was not initially recorded for eight crowdworkers. Reaching out on Prolific to get them to return to take a follow-up study (compensated at \$10 and taking 15 minutes) that included only the Calvi/Mini-VLAT portion of the study resulted in four completions. The other four were removed from the analysis, leaving us with 42 crowdworkers, for a total of 80 participants.

\begin{figure*}[t]
    \centering
    \includegraphics[width=0.9\linewidth]{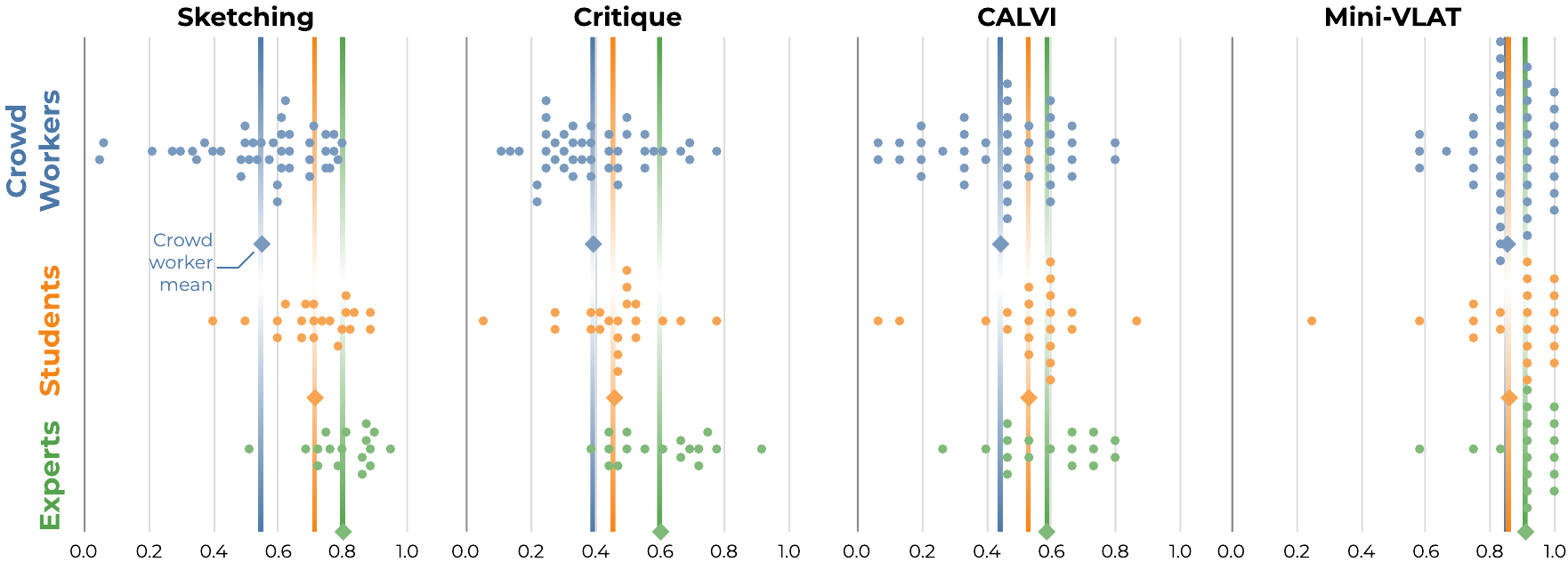}
        % \includegraphics[width=\linewidth]{figures/distributionDotplot.png}
    % \    \includegraphics[width=\linewidth]{figures/raincloudsByGroup.png}
    \caption{\textbf{Sketching and critique scores strongly differentiate between experts, students and crowdworkers.} Raw scores of participants for the four different assessments used in our study, with group means for each assessment marked. 
    % \zc{I think id just remove any CIs from here and display means, or at minimum only display the CIs of the actual average itself but, that might be confusing that were doing it differently in the other one} \mk{indicate the CI is on the difference on the direct label in the chart so people don't miss it} 
    % \am{draft pending fixes to calculation} \al{Should be Sketching, not drawing.} \am{i thought we were dropping this one?}
    }
    \label{fig:res_scores}
    \vspace{-2em}
\end{figure*}

\revised{Crowd workers had a median completion time of 58 minutes, and were paid \$20.}
Students had a median completion time of 59 minutes.
\revised{We initially advertised to pay students \$20 and enrolled a few at that rate, but had to increase compensation to \$50 due to difficulties recruiting sufficient numbers of students. Students who had previously agreed to participate at \$20 were also paid \$50.} 
Researchers had a median completion time of 74 minutes and were paid \$20 in compensation.

\revised{Nine crowd workers indicated having taken a visualization course before, and two indicated that they were currently pursuing or had a PhD in visualization. These participants were not moved into other conditions due to the difficulty confirming experience levels of crowdworkers, and the tendency of crowdworkers to overstate past experience~\cite{bell_fraud_2023}.}

\parahead{Grading}
Participant performance was scored using the rubrics described in \secref{sec:rubrics}. \revised{Before the study, there was an alignment phase during which the first and second author iteratively met and coded to develop and test the rubric on pilot data. After multiple iterations, the first and second author graded a final pilot of 23 participants' critiques (69 tasks) and 3 participants' sketches (15 tasks) using our final rubrics. Inter-rater reliability, calculated using quadratic-weighted Cohen's $\kappa$, was $\kappa=0.81$ for critique and $\kappa= 0.76$ for sketching, indicating substantial agreement in both cases~\cite{landis_measurement_1977-1}. The first author coded the data drawn from the main experiment. Answers that were ambiguous or otherwise difficult to grade and were not seen during pilots were surfaced to all co-authors in meetings to discuss and resolve.} For instance, the team discussed whether or not a drawing of a table identical to the input dataset would be a reasonable response in the sketching condition. The team agreed that a sketch of a table that was accompanied by reasoning in the transcript (\eg{} one participant argued that a table is a reasonable visualization for data of this size, saying \qt{The challenges are I don't see a better way to display this. It's quite easy to read and understand as it is.}) would receive points, while the same table without commentary would not. Other examples of tables that included a visual encoding, such as LineUp~\cite{gratzl_lineup_2013} style tables including bar charts for one or more of the values, received good scores.

\section{Results}

We avoid exposing expert and student data and replays due to the recognizability of many of our participants' voices to other members of the visualization community. Results and analysis for crowd workers are available in supplemental materials. When possible, quotes are accompanied by links to participant replays.

\figref{fig:res_scores} shows raw scores of the results faceted by the three populations. Mean scores show Sketching and Critique have roughly normal distributions and strongly differentiate between experts, students, and crowdworkers. We utilize Bayesian IRT to further investigate results.

\begin{figure*}[t]
    \centering
    \includegraphics[width=\linewidth]{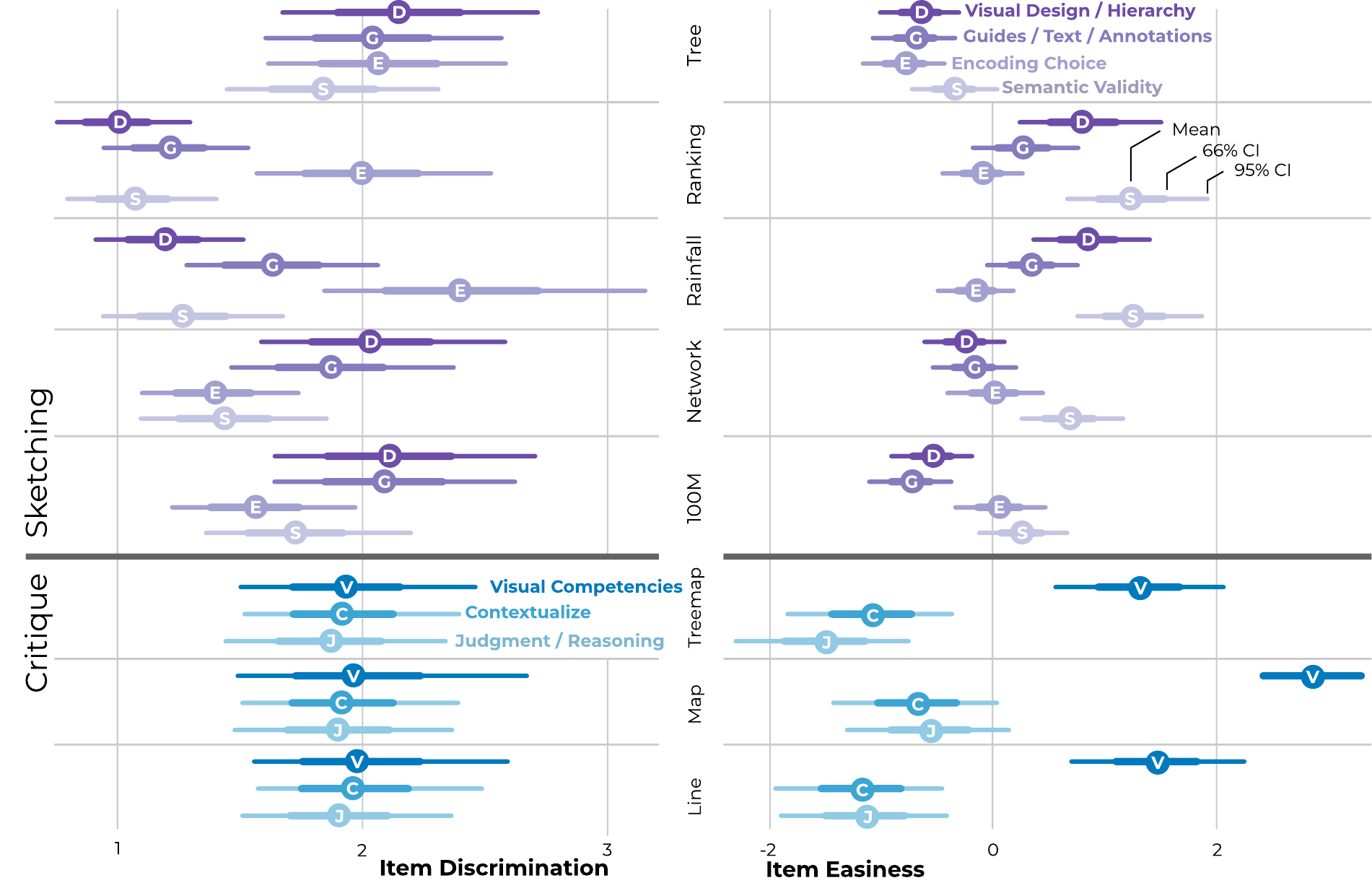}
    \caption{\textbf{Sketching tasks show similar discriminability as Critique, with a wider range. Tree, 100M, and all Critique tasks show strong discriminability, while Ranking and Rainfall tasks show slightly lower discriminability.} Item discriminability calculated from a 2PL IRT model for every question and rubric item in Sketching and Critique assessments. Each rubric item displays the mean and 66\% / 95\% credible intervals (CI) of its discrimination. High item discrimination indicates a question successfully differentiates between participants with varying ability levels. 
    }
    \label{fig:res_item-discrimination}
        \vspace{-2em}
\end{figure*}

\revised{\rqq{Will Sketching and Critique assessments be able to discriminate participants with different skills?}}
Similar to other visualization literacy assessments\cite{ge_calvi_2023, ge_avec_2025}, we utilize Item Response Theory~\cite{demars_item_2010} to assess the discriminability of our assessment questions. We selected a 2PL IRT model for all four assessments, which captures \textbf{item easiness} and \textbf{item discrimination}. We also use these models to estimate individuals' abilities on each test. Typically, 2PL IRT models utilize logistic regression, including our own models for CALVI and Mini-VLAT. However, unlike CALVI and Mini-VLAT, the Critique and Sketching assessments are graded on an ordinal scale, requiring us to instead utilize an ordinal regression model. We utilized the brms R package~\cite{burkner_bayesian_2021} to implement the models. All final models were run with 4 chains of 20,000 iterations. 10,000 warmup iterations were discarded, and the final sample was thinned by 5, resulting in 8,000 draws. The minimum bulk effective sample size between all models was 5,182, and the minimum tail effective sample size is 5,950.

Item discrimination for each part of our rubric and each item in sketching and critique can be seen in \figref{fig:res_item-discrimination}. Median item discrimination for critique items ranges from 1.86 to 1.97 for an average of 1.92. Item easiness ranges between -1.50 and 2.85, with \textit{visual competencies} being significantly easier than other rubric items for all three tasks. Sketching has median item discrimination ranging from 1.07 to 2.23 for an average of 1.52, and easiness ranging from -0.80 to 1.28. Specific ordering of discriminability and easiness for sketching rubric items varies between tasks. The \textbf{tree}, \textbf{100M}, and \textbf{network} tasks show high discriminability, while the \textbf{ranking} and \textbf{rain} tasks have relatively weak discriminability.

\begin{figure}[t]
    \centering
    \includegraphics[width=\linewidth]{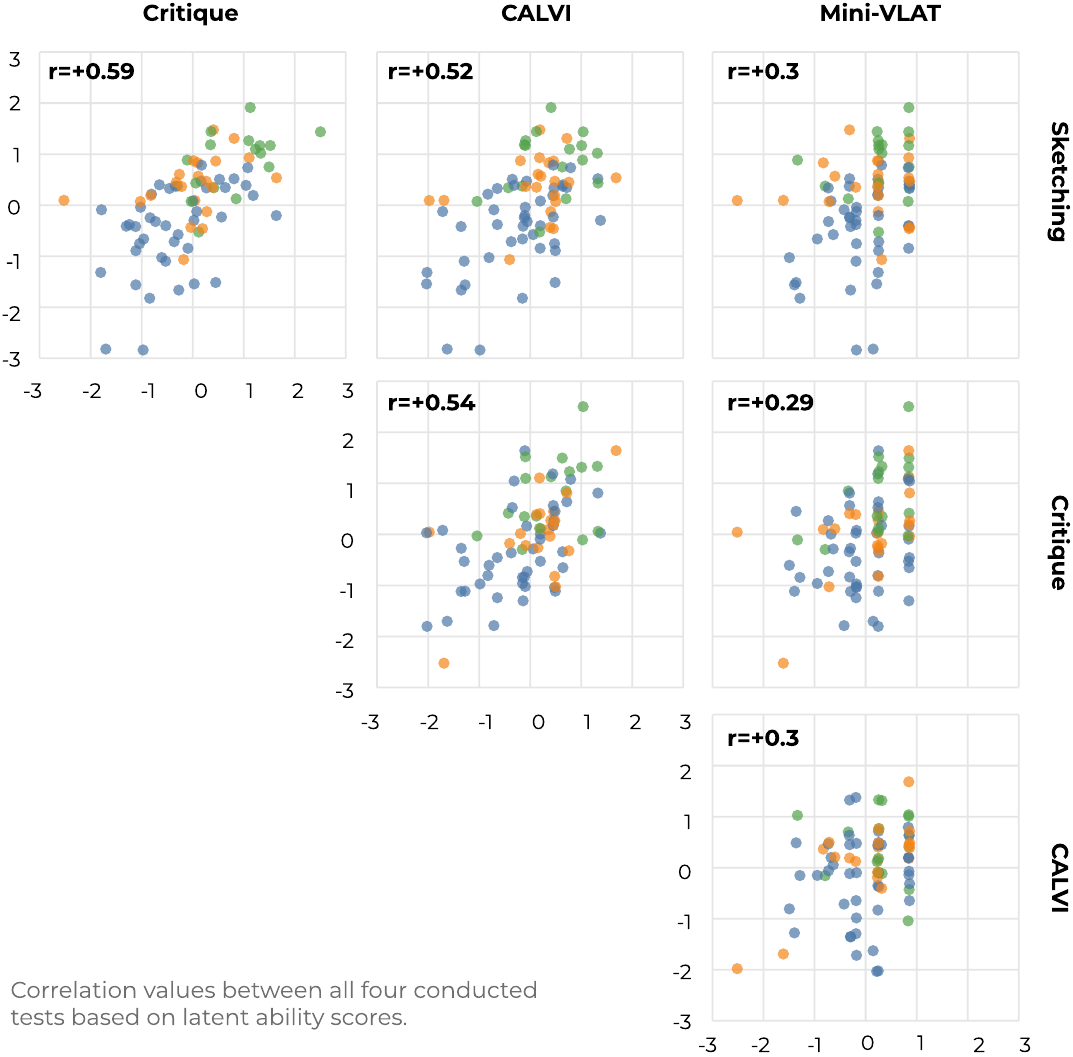}
    \caption{\textbf{Sketching, critique, and CALVI show moderate correlations with each other, while Mini-VLAT correlates weakly.} 
    % Correlation values between all four conducted tests based on raw scores.
    }
    \label{fig:correlations}
    \vspace{-2em}
\end{figure}

\begin{figure*}
    \centering
    \includegraphics[width=\linewidth]{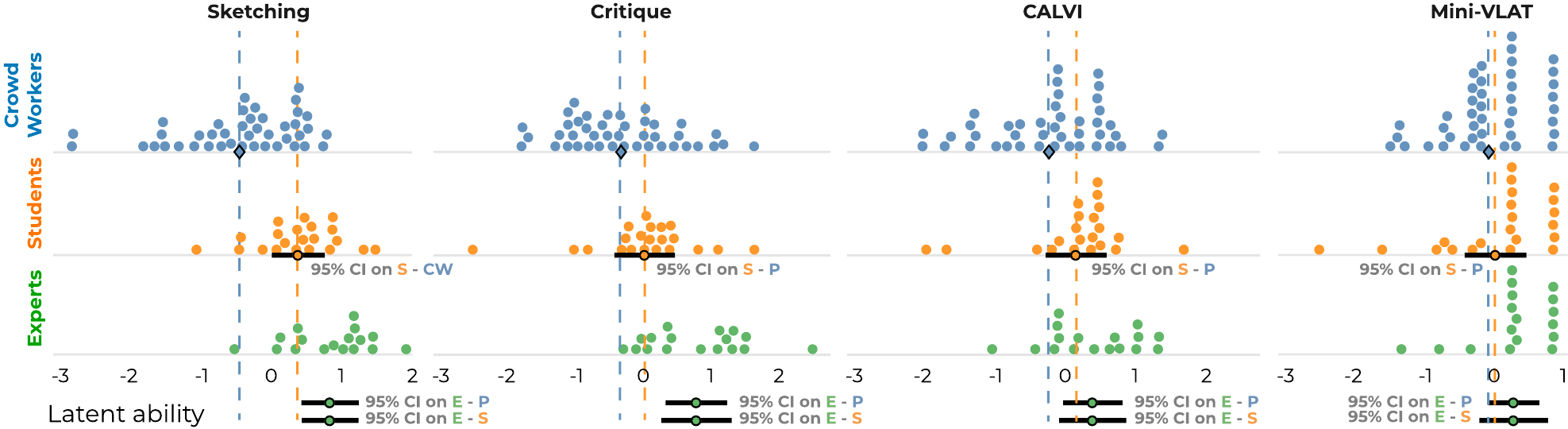}
        % \includegraphics[width=\linewidth]{figures/latentDistribution.pdf}
        % \includegraphics[width=\linewidth]{figures/distributionDotplot.png}
    % \    \includegraphics[width=\linewidth]{figures/raincloudsByGroup.png}
    \caption{\textbf{Latent ability scores derived from Bayesian IRT confirm sketching and critique differentiate between visualization expertise levels better than CALVI and Mini-VLAT.} 
    Distribution of participants' latent abilities by test and group (with corresponding means marked). Pairwise comparisons are made between each group using Welch's two-sample t-test and displayed below the Students and Experts groups.
    If a test confidence interval does not overlap with the associated mean, then the test is significant with p < 0.05.
    }
    \label{fig:latent-distribution}
    \vspace{-2em}
\end{figure*}

\rqq{Will Sketching and Critique assessments be able to differentiate between groups with varying visualization expertise?}
A distribution of participants' normalized scores on each assessment, along with the means of each group, can be seen in \figref{fig:res_scores}. To compare how well different assessments differentiate between groups, we used the estimated abilities of individuals on each test calculated from the Bayesian IRT models and ran Welch's t-test on each pair. Results can be seen in \figref{fig:latent-distribution}. We report results as (Mean, [95\% CI]), which can be interpreted as significant with $p<.05$ if the CI does not overlap zero. 

\revised{Sketching demonstrates significant differences in all pairwise comparisons, with experts and crowdworkers (1.29[0.88, 1.70]), students and crowdworkers (0.84 [0.46, 1.21]), and experts and students (.45 [.05, 0.86]) all being differentiable (also see \autoref{fig:res_scores}). Experts and students show the least differentiation of the sketching pairs, with a CI close to overlapping zero.} Critique also had significant differences between experts and crowdworkers (1.10 [0.63, 1.56]) as well as experts and students (0.74 [0.22, 1.26]), but not between students and crowd workers (0.39 [-0.05, 0.82]). As previous work has found\cite{chang_tell_2026, hedayati_what_2025}, Mini-VLAT shows a strong ceiling effect, both in the normalized scores (most of which are > 0.8) (\figref{fig:res_scores}) and in IRT-estimated latent ability (\figref{fig:res_item-discrimination}. As a result, there are no significant pairwise differences between groups, although average scores and ability levels do slightly increase with expertise levels. CALVI does not show signs of a ceiling effect, and shows significant differences between experts and crowdworkers (.62 [0.20,1.05]). Overall, both Sketching and Critique display noticeably more sensitivity between participant groups than CALVI or Mini-VLAT, demonstrating the utility of our assessments.

\rqq{How correlated are the Sketching and Critique assessments with each other and with CALVI/Mini-VLAT?} 
We calculate correlations between assessments using latent ability estimates created by our IRT models and calculating Pearson r values on both the latent ability and normalized raw scores. Correlation values based on ability estimates can be seen in \figref{fig:correlations}. \revised{Sketching, Critique, and CALVI all show moderate correlation with each other (estimated ability $r=0.52-0.59$, raw $r=0.53 - 0.57$), with sketching and critique having the highest correlation (ability $r=0.59$, raw $r=0.57$). Mini-VLAT shows weak correlation with the three other tests (ability $r=0.30 - 0.39$, raw $r=0.26 - 0.33$).}

% \begin{figure}[h!]
%     \centering
%     \includegraphics[width=\linewidth]{figures/inline-pcp.png}
%     % \caption{}
%     % \label{fig:placeholder}
% \end{figure}

% \

\subsection{Qualitative Analysis}

% Although we do not perform the kind of deep, qualitative analysis which our data deserves, while grading we did identify a few of the major differences between high and low scoring participants.
 \revised{As the aim of this work is to investigate the feasibility of our assessments, we stop short of a thorough qualitative analysis of the drawing or critique process, such as conducted by Walny \etals{}{}\cite{walny_exploratory_2015} analysis of visualization design sketches, although we believe future qualitative analysis of this or other datasets would be worthwhile. Nonetheless, in the process of grading we identified a variety of qualitative differences between high and low scoring participants worthy of future inspection.}

Sketching shows especially strong differentiation between crowdworkers and both students and experts. We observed crowdworkers frequently expressing difficulty and frustration during the sketching assessment. One participant, while completing the rainfall task, said, \ourhref{https://vdl.sci.utah.edu/visLiteracyStudy/literacy-prolific/WHUwRzhDeHVFMUh5b2VGSzNFVDhMQT09?participantId=6608914ed1513b14fb827ca4} {\qt{So easy to make bar charts and everything in Excel. When it comes to sketching it from your own hand, it's becoming a nightmare.}} Another participant got particularly upset, saying \ourhref{https://vdl.sci.utah.edu/visLiteracyStudy/literacy-prolific/Mmt5YlgrUmxYOGs0OWwxL0hxYzlyZz09?participantId=6998abd731d2ec50d067d5bd}{\qt{This is ridiculous. How am I supposed to sketch it manually without the table [expletive] processing software?}} Individuals without formal visualization training may have little experience creating visualizations without aid from systems such as Excel. Using such tools trains participants to accomplish visual mappings, but abstracts away other parts of the construction process, such as choosing appropriate base encodings and applying data transformations, leaving many participants frustrated when asked to do them during our study. 

Choosing appropriate base encodings was challenging for many participants. 
% such as the \textbf{tree} task,
For some tasks, some of the more appropriate chart forms are not well known---\eg{} for the \textbf{tree} task forms like treemaps or sunbursts would be most appropriate---and so participants without substantial knowledge would be unlikely to employ them.
% It is understandable then why a participant may not choose them.
Interestingly, many participants seemed to have a favorite chart type that they tried to fit all data types into. One participant explicitly said \ourhref{https://vdl.sci.utah.edu/visLiteracyStudy/literacy-prolific/UGczdnZrNnV5a1d4aitzaCtCNVRlZz09?participantId=5fbf94f2891c55203896f7dc}{\qt{I think I only know bar charts for some reason. For everything, I just draw bar charts.} }
Bar charts were especially common, even for the \textbf{rainfall} task where a simple line chart would be superior.

Experts frequently applied significant thought and detail in their sketches. While grading pilots, we discussed at length how critical to be of small but important visual design decisions in sketches that may differentiate between good and great charts. 
One prominent example we discussed is edge crossings in node-link diagrams. 
To maintain a high ceiling in grading, we decided to consider these details (\eg{} edge crossings). 
During the full study, multiple participants indicated intentionally avoiding edge crossings. 
One expert noted \qt{So if I put Basil in the middle, it would---the nodes would look like it's not centered. But with Clara, since she has connections to basically everyone, I put her in the middle.} 
Our high assessment ceiling allowed us to capture expertise that few participants, even among experts, demonstrated.

% \am{bad intro, show don't tell}Critique revealed a number of interesting patterns and differences between participants of varying expertise. 
Critique was similarly rich.
% Most notably, 
Crowd workers tended to produce what we called \textit{shallow} critiques, or critiques that only discussed classical rules of visualizations taught in grade school---charts must have a title, charts must have a legend, and so on. 
For example, in the \textbf{line chart} critique, many participants expressed a desire for a legend for the color encodings on each line, despite the existing direct labeling approach being an improvement in the eyes of many experts. Crowd workers also frequently made snap decisions about graphs, exemplifying anchoring effects that have been encountered in previous studies~\cite{peck_data_2019}.

Many poorly scoring participants anchored themselves to an immediate opinion, then worked backward to try and justify that opinion, making it unlikely they would produce a good critique.

% \section{AI Grading}

\section{Discussion}

Visualization literacy has typically been assessed through multiple-choice tests, as in VLAT or CALVI. Yet, as Varona \etal{}~\cite{varona_state_2025} and Leon \etal{}\cite{leon_multiliteracy_2026} highlight, visualization literacy is composed of a range of different skills and competencies, which existing assessments only partially capture. We explore how two alternative assessment modalities, sketching and critique, can be used to assess literacy. 

\parahead{Do our Assessments Measure Visualization Construction and Critiquing Skills?}
We find that the modalities investigated succeed at differentiating between participant groups. Given that sketching and critique are commonly learned through formal visualization training, observing that our assessments differentiate between experts and non-experts better than past assessments is a promising sign of the tests' content validity (how well an assessment measures what it is intended to measure). As our goal for this work is not focused on validation of our assessments, we forgo more formal validation efforts, such as utilizing external experts to calculate a content validity index for each question (as Ge \etal{}~\cite{ge_avec_2025, ge_calvi_2023} do). However, Bayesian IRT analysis of our questions reveals that both assessments have reasonable item discrimination for all rubric items, signifying the rubrics described in \secref{sec:rubrics} successfully discriminate between participants. 

% \am{something about sketching?}\al{yes!}

\parahead{Item Discriminability}
Our sketching items show a wide range of discriminability, with \textbf{ranking} and \textbf{rainfall} having relatively low discriminability, as well as a wide range of discriminability within their rubric items. This discrepancy may be due to both of these tasks having a well-known solution (line charts in both cases), leaving little room to demonstrate expertise beyond initial encoding choice. In contrast, participants who chose a node-link diagram to visualize the \textbf{network} task still had many design choices that allowed us to differentiate between participants, such as how to encode and label node size and edge weight. 
However, \textbf{ranking} and \textbf{rainfall} have higher item easiness than other tasks, indicating that these questions may be more useful when testing less experienced populations.

Our critique questions showed remarkably similar item discrimination, despite varied complexity and design issues, indicating that as a method, critique can be generalized beyond our specific charts---increasing its practical utility.
\revised{The critique questions also had similar easiness, with \textit{visual competencies} (reading a chart) being significantly easier than either \textit{contextualize} or \textit{judgment}. This gap indicates that our rubric successfully identifies \textit{contextualize} and \textit{judgment} as higher-order skills involved in critique and separates them from the baseline reading ability (\textit{visual competencies}).}

%\am{this is interesting stuff but written in a pretty dry way}
\parahead{Correlation of Tests}
Analysis of correlation between assessments reveals that Sketching, Critique, and CALVI show moderate correlation with each other (see Figure~\ref{fig:correlations})---with sketching and critique being most related, while Mini-VLAT is less correlated to the other three. The discrepancy between Mini-VLAT and the other three tests may indicate that Mini-VLAT tests a separate aspect of literacy, or that the ceiling effect of Mini-VLAT within our sample artificially limits correlation. 

Echoing Chang \etal{}~\cite{chang_tell_2026}, we find that while all of our literacy assessments are weakly or moderately correlated, they show clear signs of the distinct skills each is intended to measure. This difference is by design --- Sketching and Critique both target higher-order skills, but from different perspectives, with Critique focusing on reading, and Sketching on construction. None of our correlation values are higher than 0.6, matching similar research on text literacy that has found reading and writing assessments only have correlation values of around 0.5 across a variety of instruments~\cite{fitzgerald_reading_2000}. 

Within individuals, there are significant differences in performance across assessments that expose meaningful differences in skill. One expert participant, for instance, scored extremely well on both sketching (Normalized raw score of 0.86) and critique (0.78), but performed middle of the pack on CALVI(0.46) and Mini-VLAT(0.92). In comparison, one crowd worker who scored slightly better than the expert on CALVI (0.60) and identically on Mini-VLAT(0.92) performed much worse on sketching(0.50) and critique(0.25). Examples such as this, which are common, exemplify the different aspects of visualization literacy that each assessment captures, as well as the distinct skill profiles that each individual possesses.

\parahead{Participant Group Differences}
One prominent difference between experts and the rest of our population was a lack of shared visualization terminology. 
Many participants, especially crowdworkers, correctly understood visual encodings but described them in unique ways. For instance, one participant characterized confidence intervals around projections in the \textbf{line chart} as \ourhref{https://vdl.sci.utah.edu/visLiteracyStudy/literacy-prolific/WXo1cnRDTHVZaElFbTdmMGtjU252QT09?participantId=67b6897f104093ebe6247f04}{\qt{lighter colored \ldots auras around the dotted lines in the projected years}}. 
One participant during piloting ambiguously critiqued the \textbf{treemap} asserting that it \qt{should be a chart}. Whether they were referring to the 3D aspect, or the infographic-like appearance, or simply classifying unknown visualization types as not charts, is unclear. This ambiguity can make grading slower and more difficult, and may be less amenable to automatic grading approaches. 

\parahead{Ceiling Effects of VLAT}
VLAT~\cite{lee_vlat_2017} and related tests, such as Mini-VLAT~\cite{lee_vlat_2017} and A-VLAT~\cite{cui_adaptive_2024}, have long been a default means for testing visualization literacy. However, in our study, we observed ceiling effects in Mini-VLAT (echoing Chang \etal{}~\cite{chang_tell_2026} and Hedayati \etal{}~\cite{hedayati_pixels_2024}), in which many participants of differing skill levels got the maximum score---making it challenging to differentiate between those individuals. Particularly surprising was that we observed this effect among crowdworkers as well. We should consider the possibility that, as Ge \etal{}~\cite{ge_autoethnography_2026} discuss, visualization literacy skills change over time, and that within the populations we studied, baseline visualization literacy skills may have increased since VLAT was first introduced in 2017. Researchers should account for this ceiling effect before using VLAT (and derivatives) in instrument selection,
% as a measurement in their own studies in the future and consider alternatives, 
in particular when recruiting anything other than novices.

\parahead{Practical Considerations for Employing Our Assessments}
When choosing to use the Sketching and Critique assessments, researchers should consider the substantial amount of time involved in both administering and grading each assessment. Each sketching question took, on average, about 6 minutes in our study, while critique questions took about 2 minutes. Grading burden is similarly skewed, with sketching questions taking noticeably more time and effort to grade. To reduce the burden on both participants and experimenters, we recommend choosing a subset of our sketching questions. As previously mentioned, \textbf{rainfall} and \textbf{ranking} are often solved the same way, using line charts, and are of similar easiness and discrimination ability, indicating that choosing one of them is likely sufficient.
%have similar looking item discrimination and easiness 
%\am{wish there was a way to see this visually} \mk{does "overlap" refer to similar easiness and discrimination on all rubric items? if so could say that instead and point to that figure}, 
%with most responses to both being line charts, . 
\revised{Per \figref{fig:res_item-discrimination} \textbf{tree}, \textbf{network} and \textbf{100m} all show similar discriminability. We suggest keeping \textbf{tree} if the target population is experts, as it covers lower easiness levels (more difficult) than other questions (making it useful for evaluating experts), and also has slightly higher discriminability than other items. Other items may be used interchangeably or depending on context, such as expected familiarity with various chart types. The critique questions show similar discrimination and easiness, and could be used interchangeably or depending on context, such as using the slightly easier \textbf{map} question if the target population is non-experts.}

Finally, one obvious way to reduce the grading burden of our assessments would be to utilize LLMs for grading. Automating grading through LLMs would drastically increase the practicality of our assessments. However, LLMs have been shown to perform worse than humans when grading open-ended responses with holistic rubrics~\cite{su_essayjudge_2025, wang_evaluating_2025}, such as our assessments. LLMs also have been shown to make racially biased decisions based on dialect~\cite{hofmann_ai_2024}, which is especially concerning for our assessments that are graded from think-aloud responses rife with dialect. Nonetheless, future work should investigate LLMs as possible solutions to the grading burdens.
% introduced by our qualitative methods.

\parahead{Variations of our Assessment Procedure}
% which can be difficult to implement and quickly grow in size. 
% 
One benefit of our approach to grading via rubrics is that, for both sketching and critique, new datasets and visualizations could be easily created, and researchers should not hesitate to experiment with different stimuli in their own versions of our assessment.

\revised{A goal of this work was to demonstrate that multi-modality would allow us to target higher-order skills that multiple choice tests are not well suited to measure}, rather than to offer an explicit reusable assessment tool. While we deliberately chose our datasets and examples (see \autoref{sec:measuring}) we did not explore alternative stimuli. This leaves open a range of questions. 
For instance, would critique questions of charts with no flaws be useful? Would sketching questions with more than four rows of data be better, or sketching questions with a specific task attached to them? 
Future work should examine such combinations, and may find considerable improvements over our current stimuli and prompts through more rigorous tests of validity.

One potential use of our assessments is applying them in conjunction with multiple-choice assessments in an adaptive approach. This might involve giving participants a simpler and shorter assessment, such as Mini-VLAT, and only use \eg{} the critique assessment after a participant has demonstrated visualization literacy above a certain bar. 
Breland \etal{}~\cite{breland_writing_1999} argue that writing assessments containing both multiple-choice and essay-style questions have improved validity over either modality alone---echoing how assessments like the GRE include both reading multiple-choice questions and writing tasks.

\section{Limitations}
\label{limitations-future-work}

\parahead{Bias in Qualitative Grading} Qualitative assessments cannot avoid the influence of biases from graders, although many assessments such as the GRE go to great lengths to reduce bias through extensive training and the combination of multiple graders \cite{breland_writing_1999}. Our own assessments are particularly susceptible to bias within the expert condition, where many participants' voices were recognizable by the first author during grading. 
This suggests that technological interventions may be useful, such as preanonymization (such as via voice masking), using text-to-speech readings of the transcript, or simply only using the transcript.    
Similarly, an inherent limitation of our approach is the subjective elements and application of our rubric. While all authors were aligned throughout the process, others may judge submissions differently or prioritize other aspects of literacy during sketching or critique. 
Future work should investigate alternate means of assessing critiques and sketches, as well as further validate our rubric on a more thorough set of stimuli. 

\parahead{Participant Sample} 
Compared to evaluations of other visualization literacy assessments, our sample size is relatively small---for instance, CALVI had 497 participants. This is due to the grading burden of our assessments and the difficulty of recruiting students and experts. \revised{To assess whether the latent ability scores from our Bayesian IRT models had similar certainty for crowdworkers (n=42), experts (n=17), and students (n=21) groups, we calculated the standard deviation of each participant's posterior latent ability and then compared the average standard deviation between groups. All four assessments showed minimal differences between participant groups. CALVI groups range between $0.55 - 0.56$, Mini-VLAT groups range between $0.65 - 0.70$, Sketching groups range between $0.29 - 0.30$, and Critique groups range between $0.36 - 0.38$. These minimal differences indicate the IRT models estimated latent ability with similar precision across participant groups. Nonetheless, the small sample sizes of our student and expert groups is a limitation of this study, and should be expanded in future work.}

\revised{Many of the students in our participant pool took a particular visualization course that used one of the examples in our assessments. We did not explicitly ask if participants had seen a chart before; however, four students mentioned having seen one chart in the critique section before, the \textit{treemap}, which was used in the course. Re-analysis of pairwise comparisons after removing the students with prior exposure resulted in minimal changes, and can be found in the appendix, indicating that such prior exposure did not substantially affect our results.}
It is also likely that experts have seen some of the examples before. 

% \parahead{Alternative Qualitative Rubrics and Assessments}\am{this could also be stirred into that grading thing above}

% \am{kinda a different thought}
% Similarly, following our results, other qualitative modalities should be investigated, especially modalities that may reveal skills that critique and sketching do not. For instance, investigating visualization construction skills using physical building blocks (as Huron \etal{}~\cite{huron_constructive_2014} explore), or through games (as Adelberger \etal{}~\cite{p_iguanodon} investigate).

\section{Conclusion and Future Work}

\revised{We have demonstrated that assessments of visualization critique and sketching are useful measures of visualization literacy that reliably distinguish between participants of varying skill levels.} While the burden of our assessments is higher than classical multiple-choice questions, they might be more appropriate for contexts where construction and critique skills are essential, such as when assessing visualization literacy before an experiment on input visualization~\cite{bressa_input_2024}, or assessing preferences related to visualization design and aesthetics~\cite{quispel_would_2014, fox_quantifying_2026}.

While developing and administering our assessment, we observed a range of interesting qualitative behavior. 
For instance, the sketching task provided deep insight into how people create visualizations, how people understand and transform tabular data, as well as what they value in a visualization. 
Similarly, critique revealed what participants view as important in a chart, and allowed for clear identification of missing competencies, such as a lack of understanding of what a log scale is (or failure to recognize the scale as a log scale). 
We suggest qualitative analysis of this or similar data would likely reveal both how people conceptualize visualizations and the shortcomings that hinder higher-order visualization literacy skills---echoing Nobre \etals{}~\cite{nobre_reading_2024} investigation of poor performance on VLAT. 

\acknowledgments{
We would like to extend our thanks to the National Science Foundation (2213756, 2213757, 2402719, 2313998, DGE-2234667), to Devin Lange and all of the members of the VDL and HAVOC labs at the University of Utah for their feedback, and to our numerous study participants.
}

% \section{Conclusion}

% \begin{figure*}
%     \centering
%     % \includegraphics[width=\linewidth]{figures/drawing-examples.pdf}
%     \includegraphics[width=\linewidth]{figures/drawing-examples.png}
%     \caption{Examples of the drawings that participants generated in each of the drawing tasks (top), as well as an example of on participant making their way through the network task.}
%     \label{fig:drawing-comics}
% \end{figure*}

\bibliographystyle{abbrv-doi-hyperref}

\bibliography{2025_literacy}

% acks: paul rosen for sharing with his class, andrew's nsf grant

\onecolumn
\appendix

\section{Appendix}

\subsection{Critique re-analysis}

In order to ensure the results in the critique assessment were not influenced by the prior-exposure of participants to the \textit{treemap} plot, we re-performed the analysis without the four students who indicated prior exposure, including a new Bayesian IRT model. As with the original analysis, we ran an ordinal regression model run with 4 chains of 20,000 iterations. 10,000 warmup iterations were discarded and the final sample was thinned by 5, resulting in 8,000 draws. the minimum bulk effective sample size is 6,498 and the minimum tail effective sample size is 6,764. 

Using the estimated abilities of individuals on the critique assessment from the Bayesian IRT model, we ran Welch's t-tests on each pair of participant groups. We report results as (Mean, [95\%CI]). Experts and crowdworkers showed almost no difference from the original analysis, and was still significantly different (1.11 [.66, 1.57], original 1.10 [0.63, 1.56]). Experts and students showed slightly more differentiation than original analysis, both of which showed statistically significant differences between experts and students (0.85 [0.31, 1.39], original 0.74 [0.22, 1.26]). Students and crowdworkers showed slightly less differentiation than original analysis, and did not show statistical significance in original or re-analysis (0.26 [-0.21, 0.73], original 0.39 [-0.05, 0.82]). As with the original analysis, the code and IRT model for the re-analysis can be found in supplemental materials.

\end{document}